\documentclass[10pt, journal, twocolumn, letterpaper]{IEEEtran}

\usepackage{booktabs} 

\usepackage{epsfig}
\usepackage{amsmath}
\usepackage{subfigure}
\usepackage{graphicx}
\usepackage{cite}
\usepackage{amssymb}
\usepackage{fixmath}
\usepackage{caption}
\usepackage{algorithmic}
\usepackage{algorithm}
\usepackage[usenames]{color}
\usepackage{dsfont}
\usepackage[super]{nth}
\usepackage{color,soul}
\usepackage{algorithmic}
\usepackage{algorithm}
\usepackage{multirow}
\usepackage{comment}
\usepackage{balance}
\usepackage{paralist}

\usepackage{dblfloatfix}
\usepackage{siunitx}

\newdimen\figrasterwd
\figrasterwd\textwidth

\usepackage[normalem]{ulem}

\begin{document}

\title{Digital Twin-Enabled Domain Adaptation for Zero-Touch UAV Networks: Survey and Challenges}

\author{Maxwell McManus$^1$, Yuqing Cui$^1$, Josh (Zhaoxi) Zhang$^1$, Jiangqi Hu$^1$, Sabarish Krishna Moorthy$^1$,\\  Zhangyu Guan$^1$, Nicholas Mastronarde$^1$,  Elizabeth Serena Bentley$^2$, Michael Medley$^2$\\
$^1$Deptartment of Electrical Engineering, University at Buffalo, Buffalo, NY 14260, USA\\
$^2$Air Force Research Laboratory (AFRL), Rome, NY 13440, USA\\
Email: \{memcmanu, yuqingcu, zhaoxizh, sk382, jiangqih, guan, nmastron\}@buffalo.edu,\\ 
\{elizabeth.bentley.3, michael.medley\}@us.af.mil
\thanks{ACKNOWLEDGMENT OF SUPPORT AND DISCLAIMER: (a) Contractor acknowledges Government's support in the publication of this paper. This material is based upon work funded in part by AFRL under AFRL Contract No. \#FA8750-20-C-1021 and \#FA8750-21-F-1012 and in part by the NSF under Grant SWIFT-2229563. (b) Any opinions, findings and conclusions or recommendations expressed in this material are those of the author(s) and do not necessarily reflect the views of AFRL.}
\thanks{Distribution A. Approved for public release: Distribution unlimited AFRL-2022-5944 on 16 Dec 2022.}
\vspace{-4mm}
}

\maketitle

\thispagestyle{empty} 



\begin{abstract}

In existing wireless networks, the control programs have been designed manually and for certain predefined scenarios. This process is complicated and error-prone, 
and the resulting control programs are not resilient to disruptive changes. Data-driven control based on Artificial Intelligence and Machine Learning (AI/ML) has been envisioned as a key technique to automate the modeling, optimization and control of complex wireless systems. However, existing AI/ML techniques rely on sufficient well-labeled data and may suffer from slow convergence and poor generalizability. In this article, focusing on digital twin-assisted wireless unmanned aerial vehicle (UAV) systems, we provide a survey of emerging techniques that can enable fast-converging data-driven control of wireless systems with enhanced generalization capability to new environments. These include SLAM-based sensing and network softwarization for digital twin construction, robust reinforcement learning and system identification for domain adaptation, and testing facility sharing and federation. The corresponding research opportunities are also discussed. 

\end{abstract}



\section{Introduction} \label{sec:intro}





Unmanned aerial vehicles (UAVs) 
have been envisioned as 
a key enabling technology 
for a wide set of new applications 
because of their unique characteristics such as 
fast deployment,
high mobility, 
on-board processing capabilities, and reduced size.
This has allowed significant progress in foundational research towards UAV-assisted communication networks, e.g., swarm UAV networks. 
Specifically, the high mobility of UAVs can be leveraged to enable dynamic network area coverage and maximize service capacity at mobile ground nodes \cite{qiang2019multiUAV}. 
Furthermore, UAV swarms can serve as MIMO-enabled self-organizing flying hotspots for terrestrial ad-hoc networks to improve spectral efficiency \cite{guan2018uavmimo}. 
In IoT networks, UAVs can be leveraged as distributed relay nodes to expand coverage area and improve quality of service (QoS) \cite{qixun2019iotuav}. 
To enable 5G and Beyond network capabilities, UAVs can provide additional computational resources for offloading and support in mobile edge computing (MEC) networks \cite{XianfuChen20}. UAV swarms are also expected to enable 5G massive MIMO (MMIMO), serving as dynamic relays to enable high-throughput communications between MMIMO base stations and ground users and minimize inter-cell interference \cite{RodriguezarXiv2018, ChandharTWC2018}. 
Additionally, future network architectures, i.e., 6G, are projected to support hybrid aerial-ground communications, in which terrestrial networks, aerial UAV networks, and satellite communications are linked hierarchically to further enhance QoS and network flexibility~\cite{WenSun2020Dynamic}.


However, while UAVs can certainly enable a new range of applications,
the challenges are multi-fold.  
%
%
First, the management of UAV-assisted networks 
needs to consider the high mobility of all connected nodes, and this
requires new resource orchestration and algorithm designs to anticipate the dynamics and requirements of each flying node and hybrid link in addition to those dynamics inherent to the networking environment~\cite{Luong2019applications}.
The situation will get even worse when jointly considering 
the newly emerging sophisticated communication techniques, such as 
heterogeneous multi-band communications \cite{Moorthy2020beamlearning},
device-to-device communication links \cite{su2022d2d}, 
integrated access and backhaul (IAB) 5G networks \cite{Alghafari2022iab}, non-orthogonal multiple access (NOMA) \cite{JianboDu21, YanyuCheng21} and spectrum coexistence \cite{GuanINFOCOM16, swarmshare}.
%
%
Moreover, in the current practice of wireless engineering, the networking environments are usually assumed to be known
at design time, and the resulting
control programs may fail when encountering unforeseen conditions.
Traditional manual network management has hindered the adoption of new techniques and the evolution of wireless networks, motivating a new paradigm
that can enable 
\textit{zero-touch} management of UAV-enabled networks, including  
planning, design and deployment, service delivery, resource management, and end-to-end optimization \cite{coronado2022zerotouch}. 


\begin{figure*}[t]
    \centering
    \includegraphics[width=0.9\textwidth]{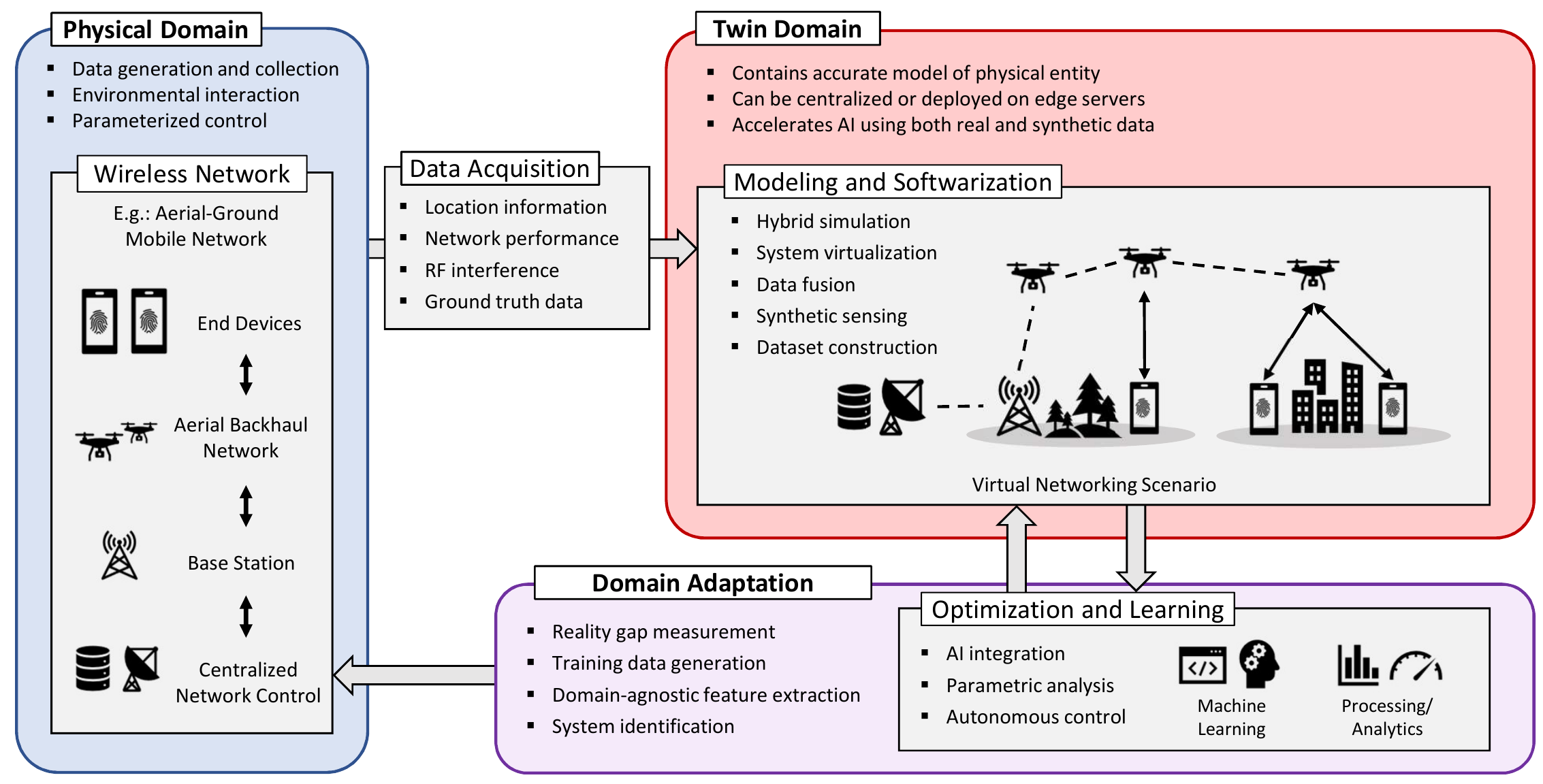}
    \caption{\label{fig:dt_overview} Top-level overview of a DT-enabled system.}
\end{figure*}

\textbf{Data-driven Approaches.}
Data-driven modeling and decision-making based on Artificial Intelligence and Machine Learning (AI/ML) are envisioned to be key enablers of zero-touch wireless network management.
In recent years, data-driven approaches based on 
AI/ML have shown great potential for automating the modeling and control of complicated wireless systems. 
Examples of recent efforts include deep learning-based edge computing for Internet of Things (IoT) \cite{HeLi18}, 
multi-label classification for user association in mm-wave networks \cite{RuiLiu2020}, 
trajectory and passive beamforming design in UAV-RIS wireless networks based on a decaying deep Q-network \cite{XiaoLiu21}, 
and network slicing for industrial IoT based on deep federated Q-learning \cite{Seifeddine21},
among others. Readers are referred to 
\cite{ChangyangNetwork20, Changyang2021Tut, Sawsan2021, AmalFeriani21} and the references therein for a good survey of the main results in this field. 

However, the primary challenges with data-driven approaches are their slow convergence rates and the limited generalization capabilities of the learned policies when faced with new environments.  
Specifically, 
the performance of ML (especially deep learning) algorithms highly relies on the availability of 
a sufficient amount of well-labeled contextual data for model training,
leading to slow convergence rates in online applications \cite{DeepMIMO2019, ChangyangNetwork20}.
Additionally, collecting training data can be too time costly and in some cases pose safety risks for hardware or network operators.
Alternatively, the models can be trained in an offline manner using data previously collected or generated by simulators \cite{SergeyLevine2020}. However, the trained models may suffer from poor robustness, i.e., it is hard for the models to generalize to new environments with different transition kernels. 

\textbf{Digital Twin-enabled Data-driven Control.}
Digital twins (DT) are envisioned as key enablers of fast-convergent and robust learning for next-generation intelligent cyber-physical systems, such as smart factories and manufacturing \cite{Jones2020Characterising, Xia2021Train, Qinglin2018}, construction, bio-engineering and automotive \cite{Ivanov2020, ROBERTOMINERVA20}, as well as wireless communication networks \cite{RuiDong19, WenSunReducing20}. 
With high-fidelity models in the virtual DT environment, 
the corresponding physical entity can be reconfigured, simulated and tested at a fraction of the cost and in a fraction of the time of deployment-based configuration testing. 
By simulating the behaviors of the physical entity in real-time, possible trajectories of a physical entity's life-cycle can be generated using physics-based simulation in order to predict events and conduct root-cause diagnosis. 
Further, the resulting data can be used to train AI/ML models 
to determine the optimal solutions for complex control problems, while the trained models can be fine-tuned 
through real-time feedback from the physical entity. 
However, while the great potential of DTs has been demonstrated in smart factory,  manufacturing and military applications \cite{Jones2020Characterising, Xia2021Train, Qinglin2018}, its adoption in wireless communication networks is still in its early phase. 

In this article, 
we aim to provide a survey of the main results of DT-enabled 
machine learning in UAV-assisted wireless networks, and discuss the research challenges and possible solutions. In existing literature, there are already 
a number of surveys and tutorials focusing on DT-enabled wireless systems
\cite{Latif21, Adil20Values, Qinglin2019Tools, ROBERTOMINERVA20, Huan20}. 
For example, in \cite{ROBERTOMINERVA20} Minerva et al. discuss the foundational properties, essential characteristics and business values of DTs focusing on IoT application domains such as digital patient, digital city and cultural heritage. The authors of \cite{Latif21} discuss DT-enabled 6G from an architectural perspective, including the key design requirements in decoupling, scalability, security and reliability as well as deployment. The enabling technologies for DTs are discussed in \cite{Qinglin2019Tools} for cognizing and controlling the physical world, DT modeling, DT data management, DT services as well as connections in DTs. In \cite{Huan20}, Nguyen et al. identify the potential benefits of DT for rolling out 5G networks, including interactive 5G emulation, 5G radio and channel emulation, and continuous validation and optimization. The application of AI/ML techniques in wireless network modeling and control has also attracted significant research attention. Readers are referred to 
\cite{Lei2020Autonomous, Luong2019DRL, Feriani2021DRL, Yichen2019Survey} and references therein for a good survey and tutorial for the main results in this field. \textit{Different from the above surveys and tutorials, in this article we discuss the challenges and enabling techniques 
for fast-convergent and robust learning in DT-assisted wireless UAV systems. }

\section{Digital Twins for Wireless Systems: A Primer
}\label{sec:frameworkDT}

Digital twins were first 
introduced in the NASA Apollo program as a \textit{``multi-physics, multi-scale probabilistic simulation"} of an object, system, or process in the physical world, which uses physical parameters, historical data, and sensor updates to provide an accurate virtual ``mirror" of the target system~\cite{Andres2019}. 
As depicted in Fig.~\ref{fig:dt_overview}, a DT system generally
consists of three 
major components: a physical entity with observable behaviors, 
a logical (or virtual) object that represents the physical entity in a simulated environment, and a bidirectional 
feedback system
between the two entities \cite{Jones2020Characterising, Qinglin2018, RobertoSaracco2019, RobertoSensing21}. 
Considering modern applications, 
DT
systems
can monitor and virtualize dynamically the behaviors of the physical systems at run-time and further aid in a zero-touch manner the decision-making in unforeseen situations based on data-driven modeling and optimization \cite{Rozanec20}. 

In order to provide accurate modeling and control decisions in spite of mathematical generalization, a DT system 
requires a bidirectional feedback loop capable of translating observed physical behaviors into a virtual model and vice versa. This behavioral translation process is termed \textit{domain adaptation}. 
To broaden the scope of this investigation, we consider a general theoretical definition of a DT with three 
critical
elements: 
the physical domain, the twin domain, and domain adaptation.
We discuss each element in the context of wireless networks with flying base stations as an example to motivate 
the application of DTs for next-generation wireless network optimization.
A survey of more general DT use cases envisioned to support 6G network capabilities such as high-density deployment configuration and reflective intelligent surface-enabled terahertz communications can be found in \cite{Kuruvatti20226GDTsurvey}.

\textbf{Physical Domain.} The physical domain is also called the \textit{target domain}. 
This domain encompasses all scenario- and application-specific aspects of the system,
such as basic network functionalities, mobility controls, physical entities, and other features of the deployment environment. In general, data acquisition is handled in the physical domain and uploaded to the twin domain in real-time or stored as a dataset for later use, which we will discuss further in Section~\ref{sec:data}. 
Considering the example of UAV-assisted networking, the physical domain would include UAV hardware, software, and communication systems used to realize the aerial base station capabilities, all comparable elements of network end-devices, as well as environmental and geographical features, such as wind speed, RF interference, and blockages, that constrain 
the UAVs' flight patterns and impact network coverage and performance.

\textbf{Twin Domain.} The twin domain, aka \textit{source domain}, encompasses all exogenous elements of the DT system 
that
are designed to accelerate optimization of physical domain applications, facilitate machine learning applications without impact on real-time system operation, or otherwise improve system performance over what is achievable in a deployment that is solely in the physical domain. 
In general, this will include virtualization of the target environment, synthetic data generation for policy convergence, and feedback with a domain adaptation process for effective policy transfer across the \textit{sim-to-real gap}. The \textit{sim-to-real gap} refers to the discrepancy in observable performance and behaviors between physical domain entities and their virtual counterparts. This discrepancy is typically caused by generalizations of unpredictable real-world phenomena present in the simulation. 
Experience collected by agents, especially in dynamic or time-varying environments, may only be valid temporarily, requiring continuous computation and re-optimization. 
Virtualization is a key technique for improving the flexibility and efficiency of zero-touch control for wireless networking systems \cite{coronado2022zerotouch}. 
This requires dedicated computational resources and infrastructure to support synthetic data generation and processing as well as bidirectional communication between twin and physical domain systems. 
We will discuss several methods of target system virtualization and softwarization for wireless networks Section~\ref{sec:virtualization}.

A detailed example of source domain design for a coordinated UAV swarm network is outlined in \cite{LeiLei21}. This example includes a centralized intelligence center collecting periodic updates of environment state data, and returning control directives to the deployed hardware for optimal MAC-layer configuration. 
The intelligence center contains a simulation of the deployed hardware capable of generating synthetic data analogous to physical domain experience, which is in turn used to train a deep neural network to optimize protocol parameters based on a physical domain scenario. 
With reliable communication between the twin (i.e., source) and physical (i.e., the target) domains, offloaded computation 
can facilitate accelerated convergence and practical applications, removing prohibitive resource constraints on physical domain systems.

\textbf{Domain Adaptation.} This is the process by which experience collected or generated
in one domain
is translated 
for use in 
the complementary domain. 
While the addition of resources from the source domain can be incredibly useful, communication between 
twin
and physical domains may not always be reliable.
In the case of unreliable communication between domains, the sim-to-real gap may be increased due to the lack of synchronization between physical systems and their virtual counterparts. 
In such scenarios, learning conducted in the
twin domain must be robust to the difference between dynamics in the physical domain and generalizations made in the source domain to enable effective \textit{sim-to-real} policy transfer.
The core focus of domain adaptation is to modify learning algorithms and source domain parameters to overcome these challenges, and is envisioned as the key to solving open research challenges associated with robustness and performance losses in transfer learning applications inherent to DT-enabled systems \cite{levine2018grasping}. 
In 
existing 
literature, domain adaptation for RL applications can be achieved by modifying 
observations of the source domain \cite{levine2018grasping}, simulation parameters \cite{chebotar2019sim2real}, 
or the reward function \cite{levine2021off}
of a well-defined Markov decision process (MDP).
In each of these approaches, a twin domain is constructed for rapid and efficient training, 
with the goal of minimizing interaction with the physical domain hence maximizing communications efficiency while maintaining effective transfer learning performance between domains. 
We will discuss DT testbed development to experimentally evaluate domain adaptation techniques for \textit{sim-to-real} policy transfer in wireless networks in Section~\ref{sec:scenarios}. 

\section{Data Acquisition}\label{sec:data}
In a DT system, the role of data acquisition is to generate and maintain a virtual environment using ground truth data from the physical domain. In addition to data required to build the virtual model, timely updates from the physical domain are necessary to maintain the accuracy of event prediction, trajectory modeling, and control capabilities in the twin domain. For a DT-enabled wireless network as outlined in Fig.~\ref{fig:dt_overview}, this time-sensitive information can include mobile base station and user locations, performance metrics, and changes to protocol specification such as modulation or bandwidth.  

In existing work, especially for physics-based or high-fidelity models, construction of the virtual environment is done manually and prior to 
simulation events based on expert understanding of the target environment. 
The majority of works discussed in 
Sections~\ref{sec:intro} and 
\ref{sec:frameworkDT} 
demonstrate the use of a virtual environment designed and deployed prior to execution time, and otherwise do not consider an explicit interactive construction of an environment model.
However, especially for dynamic physical environments 
such as UAV networks
\cite{LeiLei21}, the deployment environment may not be known ahead of time and the DT system must be able to generate blockage and boundary rules at execution time.

The authors of \cite{RobertoSensing21} describe the virtual environment of a DT system as a repository of environmental and system signatures.
Behavioral or physics-based modeling is of key importance to ensure accurate decision-making based on the virtual environment \cite{Huan20}, which requires efficient, reliable collection of high-fidelity environmental data.
New methods of collecting these signatures automatically are currently being investigated to accelerate the development and deployment of DT systems, especially with the help of robots, UAVs, or other technology to enable autonomous mapping and unassisted control.

In the following section, we discuss the enabling technologies for DT construction and deployment and discuss different methods of environmental data acquisition in this context.


\subsection{Enabling Technologies and Techniques}

We identify online environment virtualization 
using various data acquisition techniques 
as a key enabling technology for real-time and mission-critical applications of DT in unknown physical environments.
These techniques include LiDAR \cite{brock2021lidar,minos2018towards3d},
millimeter-wave radar \cite{Moallem2014MMW},
and
SLAM \cite{raul2017orbslam, karthik2020edgeslam},
to quickly and efficiently scan an environment and build an interactive virtual model. 
Once an environment is generated, contextual datasets can be generated using ray tracing or other simulation methodologies to accelerate optimization tasks as shown in Fig.~\ref{fig:dt_overview} \cite{DeepMIMO2019, LeiLei21}.
However, the 
automation of 
data acquisition
to enable online, on-the-fly DT construction is of critical importance to enabling DT for zero-touch networking. Online DT construction in general, to the best of our knowledge, is a challenge that remains unaddressed in existing DT literature.

\textbf{LiDAR.} In recent literature, LiDAR has demonstrated excellent promise for generating high-fidelity environmental models.
LiDAR systems detect surface points in an environment by emitting light in the form of
pulsed laser and calculating the time-of-flight based on reflections \cite{brock2021lidar}. These points are aggregated in the form of a point cloud highlighting key features in an environment, which is then converted into a representative mesh by a central controller (typically via numerous filtering and reconstruction steps). 

Currently, some visual sensor based autonomous vehicles use simplified LiDAR sensors to accurately measure the distance between the vehicle and obstacles, then fuse the LiDAR's data with other sensors' data to make a more robust map.
Some LiDAR-based autonomous vehicles will use high-accuracy LiDAR as a primary sensor to generate the ambient environment's map. In addition, most vacuum robots use a simplified LiDAR system to build the map of the user's house for path planning, collision avoidance, and localization. LiDAR is leveraged in \cite{brock2021lidar} to construct an interactive virtual model of an environment in the Unity gaming engine. The Unity engine was selected to maintain the 3D environment model due to its 
high-quality visualization and integrated physics capabilities. 
While gaming engines such as Unity are typically optimized for rendering and visualization with user interactivity, they are not generally capable of rendering dynamic meshes from data acquired in real-time.   

In general, LiDAR sensors, especially high-fidelity long-range sensors, can be very expensive, ranging from hundreds to tens of thousands of dollars \cite{minos2018towards3d}. The sensing range of high-accuracy LiDAR systems can reach more than 100 meters, with an accuracy of 1 mm. However, a high-accuracy LiDAR system is heavy (5 kg+) and expensive (\$10000+), while low-cost, simplified LiDAR systems' scanning frequencies are too low. Therefore, LiDAR may not be the best choice for large-scale aerial mapping, where the weight of sensors must be minimized. 
The authors of \cite{lin2019uavLidar} demonstrate the capability of a lightweight ($<$ 1 kg) LiDAR sensor for aerial mapping at an altitude of 10-20 meters. 

\textbf{Mm-Wave Radar.} In addition to optical measurement methods, the use of millimeter-wave (mm-wave) radar has attracted attention in recent literature as a method of mapping a physical environment based on measured backscattering and time-of-flight of emitted high-frequency RF signals. Specifically, the use of Y-band (215 GHz) radar for indoor navigation and mapping is demonstrated in \cite{Moallem2014MMW}. In this work, a portable mm-wave radar system is assembled using  commercial-off-the-shelf RF components interfaced with a vector network analyzer to collect range information. The system was mounted on a turntable to enable full rotational scanning, and a LabView interface was developed to control the radar position and data collection. Data was collected by emitting RF signals at 215 GHz with 1 GHz bandwidth at four different types of polarization - HH, HV, VH, and VV - 
for performance comparison.
The reflected signal response was measured at 220 GHz over 5 GHz bandwidth and processed using Hough transform, ghost image elimination, and false blockage elimination techniques to extract a 2D model of the local environment, providing up to 15 cm resolution when scanning in HH polarization. 
Since this method supports real-time map generation, it is considered a viable method for simultaneous localization and mapping of DT environments.

The authors of \cite{zhao2020mcube} introduce the M-Cube, an experimental software-defined millimeter-wave radio system which is constructed using a low-cost 802.11ad radio and a programmable baseband module. This system can provide full control over MIMO beamforming, providing up to 256 antenna elements across 8 reconfigurable arrays, and has been experimentally validated for both mm-wave (60 GHz) communication at up to 325 Mbps as well as mm-wave radar based on AoA estimation for object detection at a range of 1 m with 8 cm resolution. 
This system represents a very interesting enabling technology that improves accessibility of experimental mm-wave approaches, reducing the overall cost and complexity of applications and providing support for new sensing-based virtualization techniques.

\textbf{SLAM.} 
SLAM is the method of creating a feature map of an unknown physical environment while tracking the location of an agent traversing through the environment at the same time, using monocular, stereo, RGB-D, or other visual sensing methods. 
SLAM requires sensors to detect the environment's features, track specific features then calculate the shape of the physical environment, the carrier's movements, and its relative location. While SLAM systems can leverage a variety of sensor input types, including LiDAR and mm-wave sensors, visual SLAM (V-SLAM) is very popular among different SLAM techniques because it only requires an ordinary camera as the input sensor. 

\begin{figure}[t]
    \centering
    \includegraphics[width=0.48\textwidth]{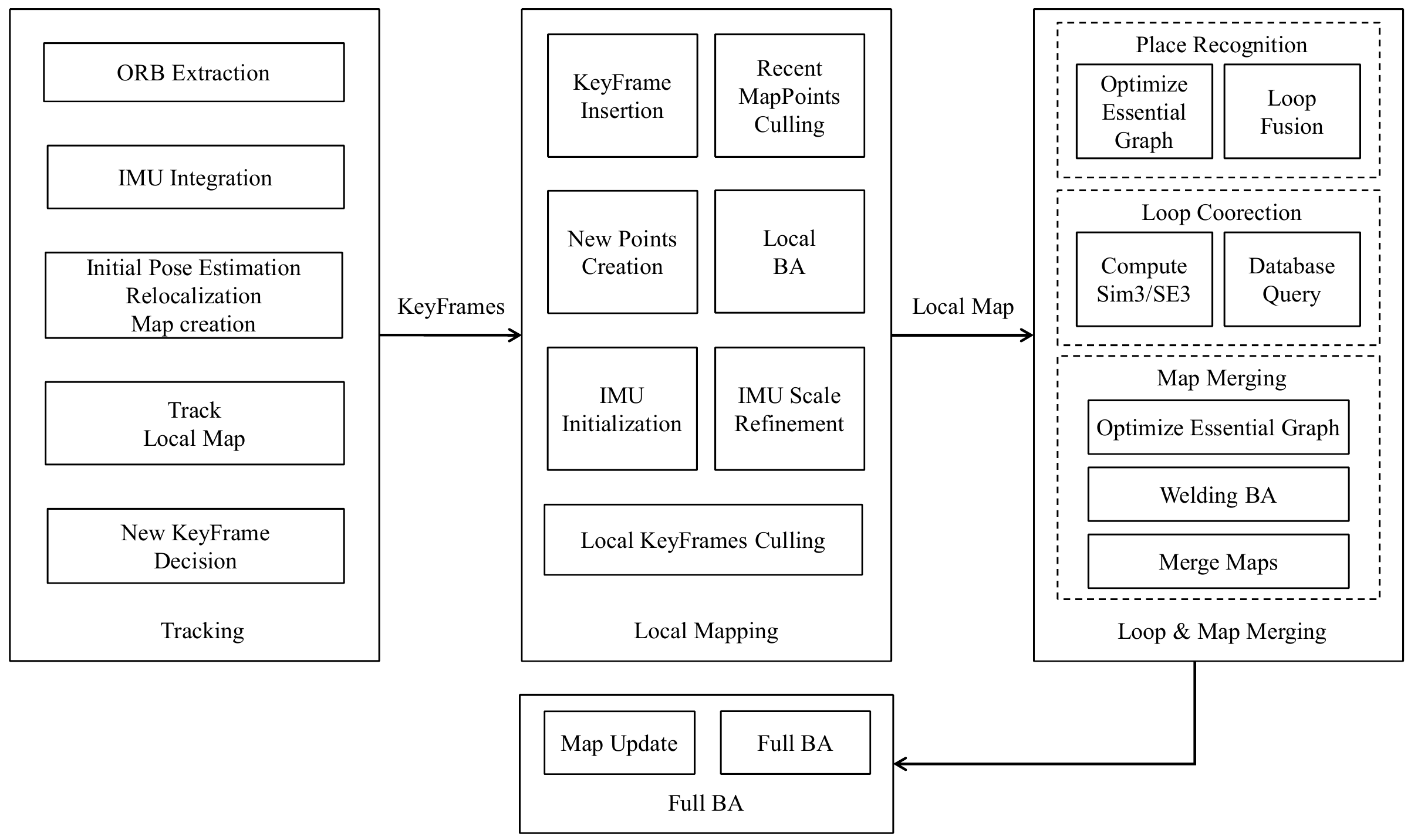}
    \caption{\label{fig:SLAM3} General diagram of ORB-SLAM3  mapping process. 
    }
\end{figure}

SLAM 
is critical to the process of autonomous virtual environment construction. The authors of \cite{campos2021orbslam3} propose ORB-SLAM3, an open-source, extensible visual SLAM framework library that supports monocular, stereo and RGB-D cameras for data collection.
In general, most SLAM algorithms are executed
following
three steps: 
tracking, local mapping, and loop closure. In the tracking phase, points in the environment are used to generate representative images of the environment called keyframes. 
During local mapping, keyframes are inserted into a local model of the environment.
Finally, the loop closure process detects and manages redundant points based on existing keyframes, integrates new information with the global model, and performs bundle adjustment (BA) to estimate camera trajectory. A more detailed overview of the processes involved in each step of this method is 
shown
in Fig.~\ref{fig:SLAM3}.
In ORB-SLAM3, the camera's image input will be fused with acceleration data from an inertial measurement unit (IMU) as shown in Fig.~\ref{fig:SLAM3}.
This can significantly improve the mapping accuracy and make the tracking continuous even if visual tracking is lost. The IMU fusion of V-SLAM, deployed in the Tracking and Local Mapping modules in Fig.~\ref{fig:SLAM3}, has been shown to surpass other state-of-the-art SLAM methods on existing datasets collected using stereo and monocular cameras. Additionally, this framework library supports data collection and processing in real time, which is critical for online DT creation and can be leveraged for simultaneous virtualization and interaction. 

The authors of \cite{minos2018towards3d} compare V-SLAM with LiDAR mapping 
using low-cost sensors for environmental mapping and mesh generation. The explored sensors include the Intel RealSense ZR300, which leverages stereo IR vision and visual-inertial odometry to perform 3D scanning and localization; the Microsoft Kinect V2, which leverages time of flight of emitted light for 3D sensing, similar to LiDAR; and the Asus ZenFone AR, which leverages a camera, a motion tracking camera, and an IR depth sensor to collect environmental signatures. These three approaches were compared to the ZEB-REVO handheld LiDAR system. 
Each sensor was mounted on an Intel Aero UAV, which navigated around a facility controlled by the native autopilot to collect environmental data. 
The environmental datasets collected by each sensor were loaded into the Unity game engine for offline visualization, observed in virtual reality using the Oculus Rift headset, and evaluated in terms of accuracy to the modeled environment and resolution of the selected hardware. It was shown that while the ZEB-REVO LiDAR sensor outperformed the other selected options, 
competent modeling performance can still be achieved for DT environment virtualization using cheaper visual sensor-based methods. Additionally, most modern passenger cars use visual SLAM for Lane Centering Control (LCC) and Adaptive Cruise Control (ACC). 

\begin{table*}
\makebox[\linewidth]{
\begin{tabular}{|c|c|c|c|c|c|c|}
\hline
\textbf{ } & \textbf{LiDAR} & \textbf{mm-Wave}  & \textbf{Monocular Camera} & \textbf{Stereo Camera} & \textbf{RGBD Camera}& \textbf{Monocular Camera-IMU} \\
\hline 
\textbf{Range}& 160m & 300m & Relative & 35m&5m&Need Further Research \\
\hline
\textbf{Accuracy}& 1.5mm & 20mm & Relative & 20mm & 3.7mm &Need Further Research\\
\hline 
\textbf{Cost}&\$10000&\$600&\$40&\$75&\$200&\$40\\
\hline
\end{tabular}
}
\caption{\label{table:sensors} Mapping metrics for SLAM systems. }
\end{table*}

\subsection{Research Opportunities and Challenges}
The adoption of these methods for DT construction provides the following key research opportunities towards enabling DT in the wireless domain.

\textit{Online DT Construction}:
The construction of a DT is broadly defined as the process by which spatial and temporal data is collected from the physical domain and used to generate a virtual environment in the twin or source domain. Online DT construction implies that data collection and virtual environment construction 
are parallel complementary processes: as the environmental data is collected, the virtual model is updated without delay. 
While \cite{Moallem2014MMW} and \cite{karthik2020edgeslam} discuss the potential for online environment construction and virtualization, the efficiency of simultaneous exploration and virtualization of an environment for use in a DT-enabled wireless simulation remains an open problem. 

Continuous SLAM is a special case of online DT construction in which 3-D environmental data is collected continuously via SLAM to update the virtual model in the twin domain. 
In general, this process will run in parallel with behavioral or analytical simulations in order to maximize spatial virtualization accuracy. 
An accurate DT simulation relies on the maintenance of the virtual model, and requires constant updates to track or model environmental dynamics in real-time. 
This is required for intelligence in the source domain to provide timely, adaptive support to agents in the physical domain. 
Continuous SLAM poses a unique challenge within the scope of online DT construction due to the amount of end-device resources, especially computational capacity and link bandwidth, required to simultaneously virtualize an environment and begin behavioral modeling. 
Furthermore, behavioral modeling, based on mobility models or historical data, may generalize too much or provide only temporarily valid solutions. 
Online, continuous generation of an interactive model presents a research opportunity which can significantly improve the state-of-the-art for next-generation wireless networks in dynamic, non-stationary environments.


\textit{Large Scale Sensing:} 
Current approaches to environment virtualization pose several limitations when considering sensor accuracy range, especially above 100 m. 
The price, range, and accuracy of several sensor types are compared in Table~\ref{table:sensors}.
The authors of \cite{minos2018towards3d, chiang2017lidar} explore the use of UAVs in expanding sensing range for environment data collection, which provides clear advantages in terms of observable area and flexibility compared to manual measurement or static sensing approaches. However, this approach poses several tradeoffs of its own: UAVs are limited in battery life, which inherently limits functional range and on-board hardware; LiDAR systems capable of collecting data at long ranges without loss of fidelity
can be prohibitively expensive; and cheaper
optical/RGB-D cameras suffer at long ranges (typically $<$ 100~m).

For LiDAR-based SLAM systems, LiDAR tracking is based on time-of-flight measurements, and LiDAR can only measure the distance between the carrier vehicle and landmarks. If there is a moving obstacle between the carrier vehicle and the landmark, LiDAR tracking may be lost. For optical-camera-based SLAM systems, camera tracking is based on angle changes. If the ambient light or viewing angle changes rapidly, then camera tracking may be lost. In either case, the relocalization process is time consuming and significantly increases computational complexity of SLAM systems.

In order to prevent tracking loss, multi-sensor fusion can be leveraged to significantly increases the robustness and mapping accuracy of the SLAM system. 
During continuous tracking, the SLAM system could use movement data of the carrier vehicle to correct the motion-caused deviation and improve tracking accuracy. 
If the tracking is lost, the SLAM system will still be able to keep updating the map with movement data acquired from the carrier vehicle GPS data or IMU unit. 

We envision one possible solution for large-scale DT construction
is monocular-IMU data fusion. 
Traditional monocular camera mapping is a low-cost, low-complexity method for large-scale, low-resolution sensing. However, a monocular camera can only measure the relative distance between objects instead of the absolute distance. Additionally, the point cloud generated by a monocular system is far less dense than other visual methods. 
To address these challenges in a UAV-based monocular-SLAM system, the camera frame can be fused with the UAV's IMU data to improve monocular mapping accuracy without additional hardware. 
Furthermore, if the UAVs use accurate GPS service, such as RTK differential GPS, 
the camera frame can also be integrated with absolute coordinates to further improve mapping accuracy by integrating collected data with geographical information system (GIS) mapping in the DT. 
However, the application of this approach to enable large-scale environment sensing and virtualization remains an open research challenge in this area.



\begin{table*}
\makebox[\linewidth]{
\begin{tabular}{|c|c|c|c|c|c|}
\hline
\textbf{Platform} & \textbf{Fidelity} & \textbf{Accessibility}  & \textbf{Type} & \textbf{Physics} & \textbf{Interface} \\
\hline 
NS-3 & High & Free, Open-source & Simulation & Single (RF) & C++, Python \\
\hline
EMANE & Low & Free, Open-source & Emulation & Multi (RF, mobility) & C++, Python, XML \\
\hline 
InSite & High & Paid, proprietary & Simulation & Single (RF) & Software GUI  \\
\hline 
Colosseum & High & Free, proprietary & Emulation & Multi (RF, mobility) & Linux VM \\
\hline
EXata & High & Paid, proprietary & Emulation & Single (RF) & Software GUI \\
\hline
UBSim & Low & Free, Open-source & Simulation & Multi (RF, mobility) & Python \\
\hline
\end{tabular}
}
\caption{\label{table:platforms} Wireless network virtualization platforms. 
}

\end{table*}

\begin{figure*}[t]
    \centering
    \includegraphics[width=0.95\textwidth]{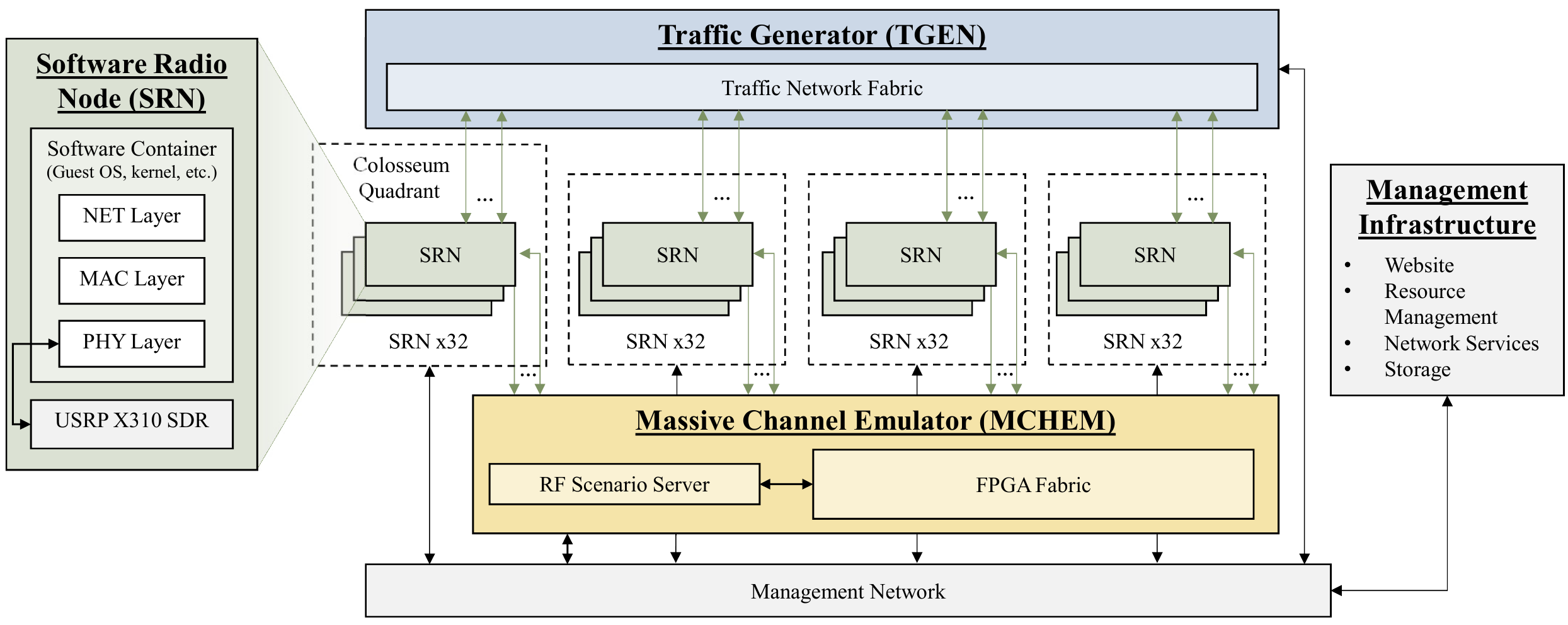}
    \caption{\label{fig:colosseum} Architecture of Colosseum \cite{bonati2021colosseum}}
\end{figure*}


\section{Network Softwarization and Virtualization}\label{sec:virtualization}

In the context of wireless networking research, the benefits of DT have attracted research attention as a key technology towards enabling highly anticipated intelligent networking tasks. 
The use of high-fidelity simulation in DT leveraging full environmental modeling for system monitoring and control is the most widely-discussed implementation observed across several industries \cite{Qinglin2018, Glaessgen2012}.

Applications of machine learning, especially 
reinforcement learning and deep learning applications, require significant amounts of time and data to generate and employ optimal control parameters for a given scenario. 
A source domain containing both simulation and optimization in a centralized intelligence center or edge server allows the synthesis of training data in place of 
experience which may be otherwise challenging or 
costly to obtain in the physical domain \cite{levine2018grasping}.
This generation of contextual data by a virtual entity,
termed ``synthetic sensing" \cite{RobertoSensing21}, is considered a key feature of high-fidelity DT. Specifically, synthetic sensing has been shown to reduce the time cost of dataset generation associated with ML applications to enable intelligent wireless network functionality, such as 
data collection and reliability, 
computational capability requirements of end-devices,
among others \cite{ChangyangNetwork20}. 
The fidelity/accuracy of synthetic data available in a DT is directly correlated to the quality and quantity of available data for a physical context \cite{Tingting2020}, as well as the capabilities of the DT platform to process this data.
Due to the growing prevalence of software-defined networking (SDN) and virtual network control, virtualization fidelity can vary widely based on the requirements of the application and the tools leveraged to create a virtual environment \cite{Qinglin2019Tools}. 
We have identified several state-of-the-art network
virtualization and softwarization platforms that have demonstrated promising synthetic sensing capabilities to address this challenge, which we will introduce later in this section. 
In addition to accurate network simulation, we identify 
several
frameworks that have been developed for establishing accurate virtual models of physical scenarios. 
While supporting experimental literature using these tools for DT development in the wireless domain
remains a key open challenge in this area, these tools offer support for the future value of DT for enabling ML applications in the wireless domain.

\subsection{State of the Art}%


We have identified several network simulators that have immediate potential for advancing research into DT for the wireless domain, including 
NS-3 \cite{ns3}, 
Colosseum \cite{colosseum}, 
EMANE \cite{emane}, 
and Remcom Wireless InSite \cite{remcom}, 
among others. Refer to Table~\ref{table:platforms} for a comparison of several key aspects of these platforms in the context of DT system design. 

\textit{NS-3:} 
NS-3 \cite{ns3} is a popular open-access, open-source network modeling tool in both industry and academia, providing high-fidelity wireless network simulation. 
NS-3 simulation is built around three major elements: nodes, which serve as basic computing device abstractions on which to run applications and install network devices; packets, which provide data flow between applications; and channels, which are used to connect nodes via installed network devices \cite{ns3}.  
All elements in a simulation are constructed from behavioral models written in C++ based on explicit protocol definition at each layer of the network stack \cite{chaudhary2012ns3compare},
with simulation control provided via C++ and Python APIs. 
Simulated network traffic can be monitored and analyzed using standard network observation software, such as Wireshark \cite{wireshark}.
This tool can be directly interfaced with radio hardware such as USRP to perform network emulation as well, improving accuracy of modeled networks.

While NS-3 can provide high-accuracy network simulation, this tool does not support explicit modeling of a physical networking environment.
Additionally, NS-3 provides very limited native infrastructure for simulation visualization and data processing and analysis, which necessitates the use of 3rd-party software for these tasks. 
While its accuracy is still limited by generalizations inherent to model-based simulation \cite{fuxjaeger2015ns3validation}, this tool is expected to play a significant role in the development of DT-enabled systems in the future due to its high fidelity, accessibility, and large community support. 

\textit{Colosseum:} 
Colosseum \cite{colosseum} is the world's largest network emulator, comprised of 256 software-defined radios (SDR) to provide a wide variety of emulated RF propagation scenarios. The architecture of Colosseum is shown in Fig.~\ref{fig:colosseum} \cite{bonati2021colosseum}. Each of the 256 SDR nodes (SRN) is made up of a software container that specifies physical, link, and network layer protocol, as well as a USRP X310 SDR which transmits data generated in the traffic generator (TGEN) based on this protocol stack. The generated signals are transmitted through an FPGA fabric in the massive channel emulator (MCHEM), which is configured to apply channel effects by emulating predefined scenarios stored on the RF scenario server. The platform is accessed, managed, and maintained using a management network connected to all constituent elements.

Similar to NS-3, it is considered an open-access tool, and provides support for a variety of different protocols including 4G/5G and IoT-type protocols with spectrum sharing. 
While there are many different predefined physical networking scenarios available, current support for custom scenarios is very limited, reducing its flexibility in the context of DT-enabled network deployments. 
We identify the need for an expansion of this framework to include user-definable networking scenarios with complete control over both network topology and communications protocol \textit{and} agent mobility and behavioral modeling. 


\textit{EMANE:} 
The Extendable Mobile Ad-Hoc Network Emulator \cite{emane}, or EMANE, is an open-source real-time framework for highly flexible simulation of mobile network systems. 
Modular network development allows for independent physical-layer modeling of each network element, providing accurate virtualization of system performance by considering signal propagation, antenna profile effects and interference sources between each emulated wireless link.
In general, each emulator instance is comprised of a physical layer model instance paired with one or more radio waveform models, which are designed in C++ and configured using XML. 
Similar to the Colosseum MCHEM, emulation instances are linked to a shared multicast channel which generates over-the-air network behaviors such as signal propagation, antenna effects, and interference. 
Additionally, EMANE provides radio waveform model plugins compatible with SDR hardware to enable shared-code emulation, which is comparable to NS-3 emulation capabilities. While EMANE is limited to emulation of PHY and MAC layers, emulation of NET layer and above protocols is typically handled in practice through integration with the Common Open Research Emulator (CORE) \cite{ahrenholz2011emanecore}. 

\textit{Wireless InSite:}
Remcom Wireless Insite \cite{remcom} is a proprietary electromagnetic (EM) propagation modeling tool for wireless networking, 
which can be leveraged for MIMO dataset generation \cite{DeepMIMO2019} via ray tracing
as discussed in Section~\ref{sec:intro}. 
The propagation behaviors are designed around several modeling theories such as 
Shooting Bouncing Ray (SBR), 
Adjacent Path Generation (APG),
and Finite Difference Time Domain (FDTD), considering environmental reflection behaviors based on the 
Uniform Theory of Diffraction (UTD) \cite{remcom2013fidelity}. 
From these models, this tool can recover significant receiver-side information such as received power, path loss, direction of arrival, delay spread, and interference estimates. 
Due to its high-fidelity modeling capabilities, this tool provides significant potential for accurate physics-based network event simulation based on signal behaviors within the propagation environment. However, this level of fidelity comes at a significant time cost. While APG with GPU can accelerate some scenarios, in general this tool will require several minutes to calculate network performance for a given deployment, preventing faster-than-real-time applications \cite{remcom2013fidelity}. 
Additionally, each scenario will need to be fully re-calculated in the case of mobile transmitter, and partially re-calculated in the case of mobile receiver, further increasing this time cost in mobile networking scenarios. 

\begin{figure*}[b]
    \centering
    \includegraphics[width=0.9\textwidth]{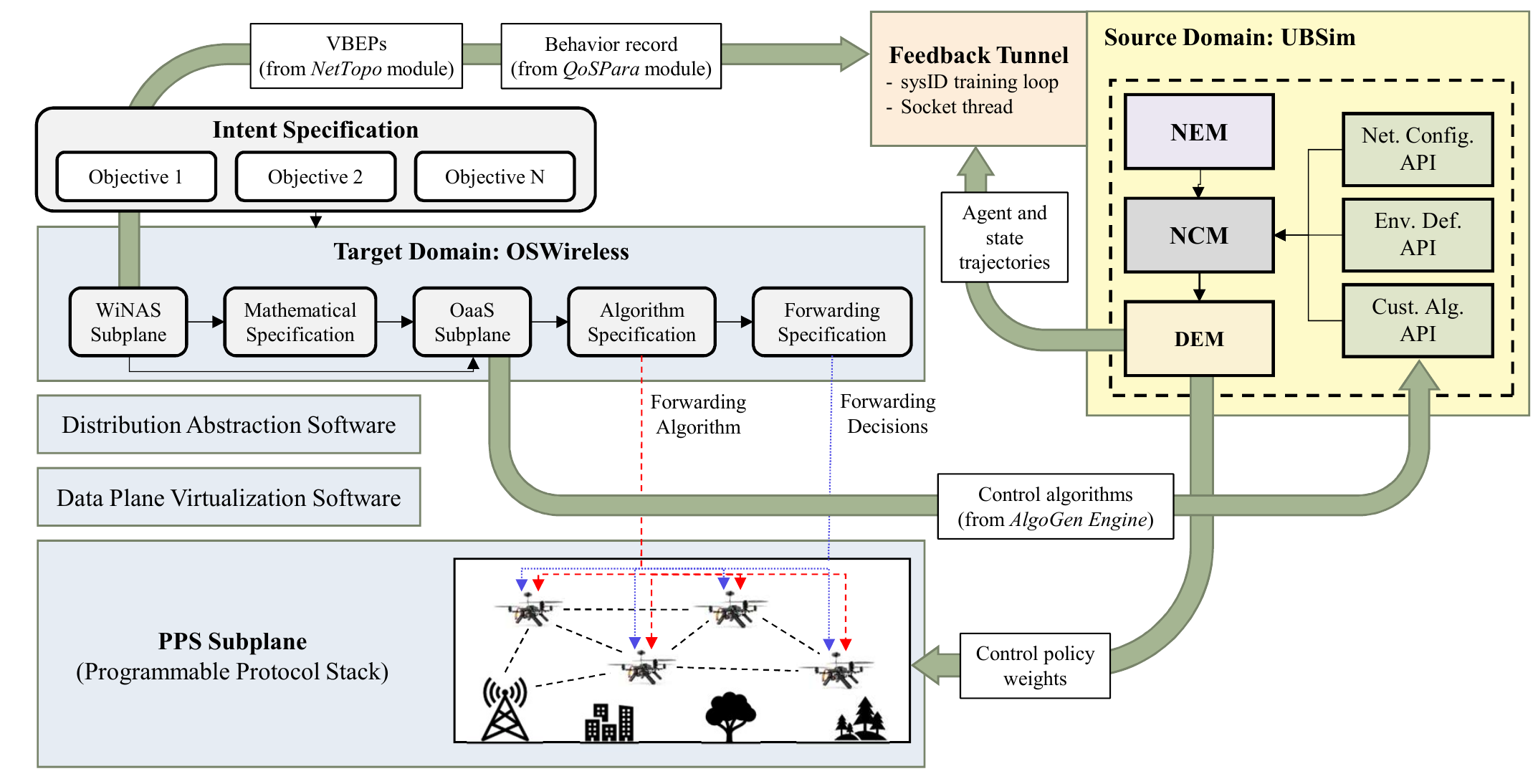}
    \caption{\label{fig:ubsim_extension} Expansion of UBSim to include OSWireless as a physical domain framework, considering accelerated control algorithm convergence and practical system identification. 
    }
\end{figure*}

\textit{ANSYS Twin Builder:}
The ANSYS Twin Builder \cite{twinbuilder} presents an open platform for DT development, with a set of built-in tools for physical and behavioral modeling of physical objects. This platform supports multiple modeling domains and languages, enabling multi-physics simulation and heterogeneous data fusion for operation \cite{Huan20}. 
Specifically, the core capabilities of Twin Builder rely on two key elements: a multi-domain systems modeler, which can simulate interactions between synchronous modeled systems based on model libraries 
such as
mechanical, hydraulic, and electronic components, logic blocks, and characterized manufacturer's components; and a multi-domain systems solver, which uses existing physics libraries for hydraulics, electronics, pneumatic systems, and thermodynamics to simulate model behaviors. 
While the platform itself is not immediately optimized for the wireless domain, its use for simulation of hardware within an IoT network, without explicit wireless network dynamics, is discussed in \cite{Qinglin2019Tools}. Twin Builder supports integration of third-party platforms as well, which implies  compatibility with other tools capable of explicitly modeling wireless network behaviors.

\textit{Spirent 5G DT:} 
The Spirent 5G Digital Twin \cite{spirent} presents a very robust platform for emulating a full end-to-end 5G network. Various 5G network elements, including independent channel emulation, virtual EPC and gNB, and full-stack end-device emulation, are virtualized with very high fidelity in order to generate cost-effective accurate behavioral analysis, enable evaluation of new security protocols, as well as other ``testing on demand" services. 
The authors of \cite{Huan20} outline several key functionalities of this platform, including wireless network automation and optimization, network slicing via SDR and network functions virtualization (NFV), and accelerated 5G network planning and validation, among others.

\textit{Pavatar:} In the context of the Internet-of-Things (IoT), intelligent online monitoring systems envisioned for smart cities, Industrial IoT (IIoT), and other next-generation IoT systems are anticipated to play a key role in supporting virtual environments constructed to support a high degree of virtualization \cite{KAZI2017}. 
A key example of the capability of high-fidelity DT supported by distributed heterogeneous sensor networks is the Pavatar system \cite{Yuan2018}. 
Pavatar collects data at multiple system layers simultaneously in order to 
conduct comprehensive sensing of all system components 
and human activities in the operation environment.
This heterogeneous data, which is in excess of 1 TB per day, is used to construct a VR representation of every system element for human interfacing, 
as well as conduct error prediction, anomaly detection, and root-cause diagnosis \cite{Yuan2018}. 
The use of different types of data sources (e.g. RF, optical, temporal) to construct a robust system virtualization, termed ``data fusion", is necessary to provide accurate simulation of physical system behaviors \cite{ROBERTOMINERVA20}. 
In \cite{Tobias2017}, emphasis is placed on detailed modeling of end-device behavior and dynamic agent-based interactions in the virtual space as an integral component of DT for distributed or decentralized networks. 

\textit{Keysight EXata:} EXata \cite{exata} is a network digital twin development and analysis tool which uses network emulation and simulation for network virtualization. The platform is based on a software virtual network (SVN) to generate each protocol layer, antenna, and device in the twin domain. This SVN is stated to be interoperable with real radio hardware and capable of interacting with real applications. Additionally, the simulation kernel leverages parallel discrete-event drivers during runtime, which can enable faster-than real-time processing necessary for improved real-time ML algorithm training and deployment.

\textit{UBSim:} 
UBSim is a custom hybrid network simulator designed for use in DT research. It is capable of simulating microwave, millimeter-wave, and terahertz-band communications in terrestrial, aerial, or hybrid aerial-ground networking scenarios 
deployed in a fully configurable physical
networking area. 
It is fully open-source and open-access, written in Python for flexibility, and ease-of-use. 
It is comprised of three core elements: the network element module,
which provides behavioral definitions of all available simulation elements; the network control module, which provides control over all deployed network elements; and the discrete event module, which schedules simulation events. 
To facilitate ease of use, three sets of APIs have been designed: the environment definition API, which coordinates all environmental features and blockages; the network configuration API, which specifies network topology and communications parameters; and the custom algorithm API, which provides templates for data-driven algorithm deployment.  
Each UBSim instance also provides a feedback tunnel, as indicated in Fig.~\ref{fig:ubsim_extension}, to enable socket communications with external software. 
In order to enable research into key technologies for comprehensive DT as outlined in Section~\ref{sec:frameworkDT}, UBSim supports parallel learning across multiple simulation instances, configurable \textit{sim-to-sim} policy transfer\footnote{Sim-to-sim policy transfer is similar to sim-to-real policy transfer, but the policy is transferred to another simulation environment instead of the real environment.} for rapid evaluation of domain adaptation algorithms such as robust learning and system identification, and 
is currently being modified to enable online DT construction using SLAM. 
UBSim has been leveraged for experiments in domain adaptation \cite{mcmanus2022sourcetotarget}, UAV network virtualization and optimization \cite{moorthy2022simsocket}, and acceleration of machine learning for wireless \cite{hu2022ubnext}, among others. While the simulation fidelity of UBSim is low, its flexibility is intended to enable integration with high-fidelity platforms such as NS-3 or RF-SITL \cite{mastronarde2022rfsitl} to enable rapid design and evaluation of DT systems through multi-fidelity, multi-physics experimentation.

\subsection{Research Opportunities}

The tools discussed in this section provide an interesting scope of customizable, potentially interoperable network simulation at varying levels of fidelity for network virtualization as introduced in Section~\ref{sec:intro}.  
However, very few tools have been accepted to be individually suitable for full DT implementation following the interdisciplinary feature set shown in Figure~\ref{fig:dt_overview}. 
Additionally, due to the lack of open-source and community support, they may not provide the level of accessibility required for rapid experimental development in this area.
The contribution of a readily available, community-oriented DT platform for the purpose of ML-based wireless network experimentation remains a significant open challenge. It is expected that a combination of these tools, combined with data fusion \cite{coronado2022zerotouch} and platform integration \cite{moorthy2022cloudraft}, can be leveraged for a widely available, high-fidelity DT toolchain optimized for use in the wireless domain. 

\textit{Multi-fidelity Simulation}: While data-driven methods can provide significant improvements to network performance and services, they can be very time-consuming and data-expensive, and may require re-training if the target environment changes over time. To balance the tradeoff between optimization time and algorithm accuracy, we identify multi-fidelity simulation as an important element of future DT-enabled wireless networking systems. 
Low-fidelity simulation can be used for time-sensitive tasks, such as rapid control decisions, by leveraging an approximation of wireless network behaviors based on observable performance statistics,
while high-fidelity models can be used to maximize network performance in stable environments, or leverage offline optimization algorithms for event prediction. 

\textit{Standardization:} Generating a standardized framework to facilitate high degrees of virtualization in the wireless domain presents many challenges, including 
simulation for highly complex and volatile networking environments \cite{Tingting2020}, 
support for dynamic, heterogeneous, and distributed network architectures \cite{Huan20}, 
and ready integration with machine learning and model-based network control \cite{Changyang2021Tut}. 
Specifically, we identify the need for an open, accessible DT framework that supports configurable network virtualization at multiple levels of fidelity. The authors of \cite{moorthy2022simsocket} demonstrate preliminary work in this area, by designing middleware between a high-fidelity UAV network virtualization platform for environmental definition with a low-fidelity network simulator for accelerated convergence of control algorithms. The continued development and distribution of such a framework would enable many contributions in this area, providing a stronger definition of the capabilities of DT technology for use in next-generation wireless networks \cite{Kuruvatti20226GDTsurvey}. 


\textit{Faster-than-real-time Optimization}: 
As part of the envisioned model for 6G network architecture, DT is expected to play a large role in the real-time or faster-than-real-time optimization of network deployments. In order to provide high-fidelity simulation -- hence accurate control directives -- protocols designed for UAV-assisted
communications will require support in virtual environments. 
Generalizable support for custom wireless protocols is possible on SDR hardware, 
and can be enabled based on integration of UBSim with OSWireless \cite{moorthy2022oswireless}. 
OSWireless is a wireless network operating system capable of decomposing operator intent to explicit network control algorithms in a zero-touch manner.  
Towards practical sim-to-real experimentation, we envision an expansion of UBSim to support integration with OSWireless, as detailed in Fig.~\ref{fig:ubsim_extension}. 
Specifically, 
OSWireless can serve as a physical domain control system to decompose operator intent into custom control algorithms. 
These control algorithms can be uploaded to UBSim, along with 
network state data
from the Wireless Network Abstraction Specification (WiNAS) Subplane, for faster-than-real-time policy training based on low-fidelity simulation. UBSim will return the optimized control policy to OSWireless for deployment on hardware nodes. To address the \textit{sim-to-real} gap, 
UBSim will leverage the WiNAS Subplane data to perform system identification, improving fidelity by tuning parameters to match simulation performance to physical domain observations.



\section{Domain Adaptation Techniques} \label{sec:adaptation}

As discussed Section~\ref{sec:virtualization}, synthetic sensing is a useful method to accelerate ML algorithm convergence for practical real-world applications \cite{ROBERTOMINERVA20, Harish20}. 
However, synthetic data is unable to fully represent all system behaviors in the physical domain, and thus introduces some inaccuracy during algorithm training. 
Many works seek to leverage high-fidelity models to improve accuracy of synthetic data \cite{Tehrani2021scenarios, Baoling20}, but processing high-fidelity behavioral models can be quite time-consuming, taking possibly several minutes to render a single environment \cite{remcom2013fidelity}. 
This trade-off between simulation fidelity and processing time may be problematic for time-critical applications. 
Domain adaptation seeks to minimize the need for this trade-off by 
improving the capability of simulation-accelerated learning frameworks to generalize from simulation in the source domain to real-world deployment in the physical domain. 

\subsection{State of the Art}

Instead of seeking a tradeoff between simulation fidelity and operation time, domain adaptation seeks to improve the generalization capability of low fidelity simulation. 
The ideal domain adaptation system seeks to minimize the importance of simulation fidelity on the accuracy of the resulting control policy in the physical domain, instead focusing on solving the contextual mismatch between domains and overcoming inherent simulation generalization to make sure the resulting control policy works. 
We introduce three representative examples of this line of research as system identification \cite{kadian2020sim2real, jiang2021simgan},
domain-agnostic feature extraction \cite{romijnders2019normalization, pan2011transfercomponent}, 
and robust learning \cite{wang2021robust, mcmanus2022sourcetotarget, khodabandeh2019robustDA}. 
In system identification, source domain simulation parameters are adapted based on feedback from the physical domain to improve behavioral accuracy. 
In domain-agnostic feature extraction, contextual features are identified from low-level data to generate a shared observation space across domains. 
In robust learning, the gap between source and physical domains is estimated through feedback and considered during training in the source domain. Readers are referred to \cite{Chen2020Robust} and references therein for a survey of robust
reinforcement learning specifically and \cite{kouw2021daReview, wilson2020daSurvey} for other general domain adaptation techniques.

\begin{figure}[t]
    \centering
    \includegraphics[width=0.4\textwidth]{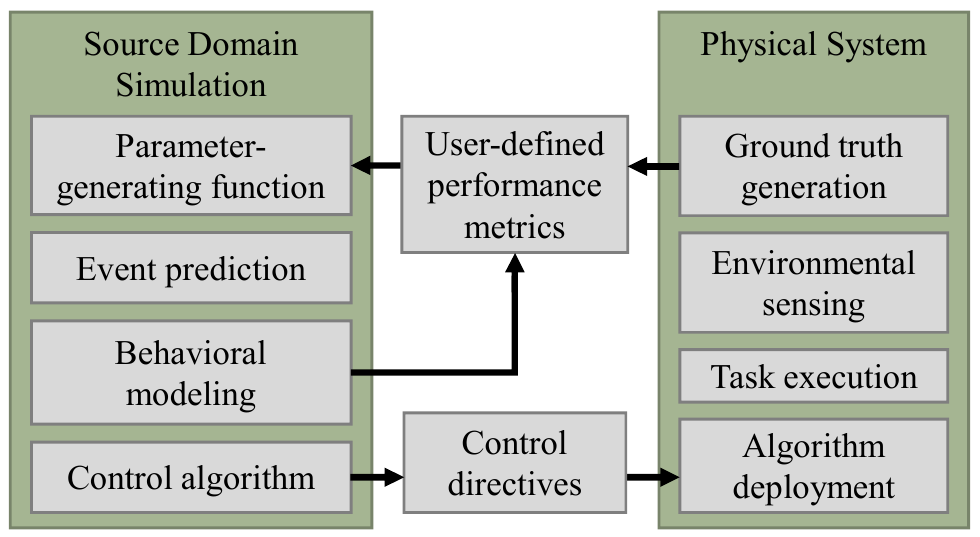}
    \caption{\label{fig:sysid_general} General outline of system identification. 
    } \vspace{-7mm}
\end{figure}

\textit{System Identification.} System identification is the most established approach to domain adaptation in existing literature \cite{levine2021off}.
In practice, system identification seeks to iteratively improve behavioral parameters in a source domain simulation based on feedback from the physical domain. An outline of the general premise of system identification is depicted in Fig.~\ref{fig:sysid_general}.
While there are many application-specific variants of system identification, this approach faces some challenges in general. 
Primarily, a significant amount of feedback is required from the target system to validate source domain performance. 
Additionally, some virtualization platforms such as Colosseum and InSite introduced in Section~\ref{sec:virtualization} may not be fully open-source or parameterized to support system identification. Such simulation platforms that are configurable and offer full control of behavioral parameters are termed \textit{hybrid simulators}, due to their analytical and behavioral modeling capabilities. 

The authors of \cite{kadian2020sim2real} explore sim-to-real transfer learning using robot navigation tasks, in which it was noticed that learners in the source domain are capable of exploiting a given simulation to perform tasks beyond capabilities in the real world. By modifying the simulator based on this feedback, the source-to-target gap was reduced and the similarity between simulated and real robot performance was improved. Similar to system identification, the use of domain-agnostic features can be used to enable domain adaptation. Instead of converging simulation parameters to maximize behavioral similarity, source domain observations can be adapted to extract high-level features common to both the source and physical domains, minimizing the impact of domain-specific dynamics on transfer learning performance. The authors of \cite{bousmalis2018domainadaptation} and \cite{romijnders2019normalization} demonstrate this practice using pixel-level observation adaptation for policy transfer in tasks related to computer vision. 

\textit{Domain-Agnostic Feature Extraction.} This method seeks to reduce the effect of the source-to-target gap by finding commonality between each domain. Specifically, this approach seeks to align behavioral inferences made in each domain through extraction of high- or low-level features that can minimize domain-specific phenomena.
For example, the approach to semantic image segmentation outlined in \cite{romijnders2019normalization} leverages pixel-level segment representation and classification to improve unsupervised adversarial domain adaptation for computer vision. The authors of \cite{pan2011transfercomponent} propose transfer component analysis, which seeks a set of features, termed transfer components, to minimize the difference in data distributions between domains. By projecting transfer components onto a shared latent space and applying standard machine learning models for classification or regression tasks.  

\textit{Robustness Mechanisms.} Robust learning 
aims to mitigate the effect of environmental perturbances, such as modeling errors, time-varying dynamics, or unreliable data, on the resulting control policy. This can be typically accomplished by applying a random or adversarial noise process to a system during policy training. 
Defined in the scope of the DT framework outlined in Fig.~\ref{fig:dt_overview}, this noise is generated by the source domain during policy training to mitigate the effect of the source-to-target gap during policy transfer \cite{wang2021robust, khodabandeh2019robustDA, mcmanus2022sourcetotarget, derman2018softrobust}.
In many cases, the noise is added in the form of training samples manually selected from worst-case scenarios or an average of potential environmental anomalies. 
This is intended to generate a policy that will provide better generalization than non-robust policies when faced with unexpected, unknown, or adversarial physical domain dynamics, at the cost of reduced maximum achievable performance. 

An effective approach to applying this policy noise to generate a robust policy is the R-contamination model \cite{wang2021robust}. Leveraged in \cite{wang2021robust} and \cite{mcmanus2022sourcetotarget} to implement model-free robust reinforcement learning, the R-contamination model is used to probabilistically alter, or ``contaminate," observations made by the agent with random or worst-case dynamics to encourage conservative policy learning. Instead of an agent following a deterministic transition kernel $p_s^a$ for a given state $s$ and action $a$, this contamination probabilistically cause an arbitrary state transition $q$, selected from an uncertainty set $\mathcal{P}$, which is comprised of all possible transitions in an environment. In order to model contaminated agent trajectories, the R-contamination model generates a subset ${\mathcal{P}}_s^a$ of $\mathcal{P}$ for 
each $s$ and $a$ pair according to ${\mathcal{P}}_s^a = (1-R){p_s^a} + R{q_s^a}, {q_s^a}\in{\Delta}_{|\mathcal{S}|}$,
where ${\Delta}_{|\mathcal{S}|}$ is the simplex of state space $\mathcal{S}$, and $R$ represents the probability of state transition according to $q$.

It is shown in \cite{mcmanus2022sourcetotarget} that the selection of random parameters from the environment can improve policy transfer performance when the source and physical domains have different transition kernels due to differences in the environment dynamics. Both \cite{wang2021robust} and \cite{mcmanus2022sourcetotarget} demonstrate the requirement for careful parameter selection prior to training, highlighting a key limitation of robust learning in the context of domain adaptation. 
While robust learning can provide very conservative policies, the authors of \cite{derman2018softrobust} propose soft-robust learning, which takes an average over the uncertainty set instead of selecting worst-case scenarios to reduce the conservative nature of the resulting policy. This yields a model capable of generalization while limiting performance degradation. 

\subsection{Research Opportunities}


\textit{Expertise Incorporated Learning}: In order to advance the use of domain adaptation for wireless networks, we consider constraint sampling reinforcement learning (CSRL) \cite{mu2022csrl} as a promising method to quantify the reality gap using domain expertise. In this way, expert knowledge of the networking environment can be integrated during the training process via sensing \cite{karthik2020edgeslam, brock2021lidar} to enable effective policy transfer in unknown environments with minimal human interaction. Sophisticated applications of this approach, especially in the wireless domain, remain an open challenge in this area.

\textit{Reality Gap}: 
The mathematical generalizations present in all simulations make sim-to-real transfer a persistent challenge due to the inherent non-linearities of natural phenomena. 
Sim-to-sim experiments can be used to estimate the domain transfer performance of new algorithms for domain adaptation, but even high-fidelity models of physical domain hardware are incapable of predicting performance exactly. 
Additionally, while synthetic sensing can help accelerate convergence of data-driven models, synthetic data will introduce inaccuracies to the learning model based on generalization \cite{jiang2021simgan}.

To overcome the reality gap between simulation results and physical system behaviors, the authors of \cite{Chen2020Robust} discuss different applications of robust learning to provide performance guarantees in the presence of environment uncertainties.
While some contributions address challenges associated with the reality gap \cite{levine2018grasping, kadian2020sim2real}, sim-to-real adaptation in the wireless domain remains an open research area.

\begin{figure*}[b]
\begin{center}
\begin{tabular}{cc}
\includegraphics[width=0.48\textwidth]{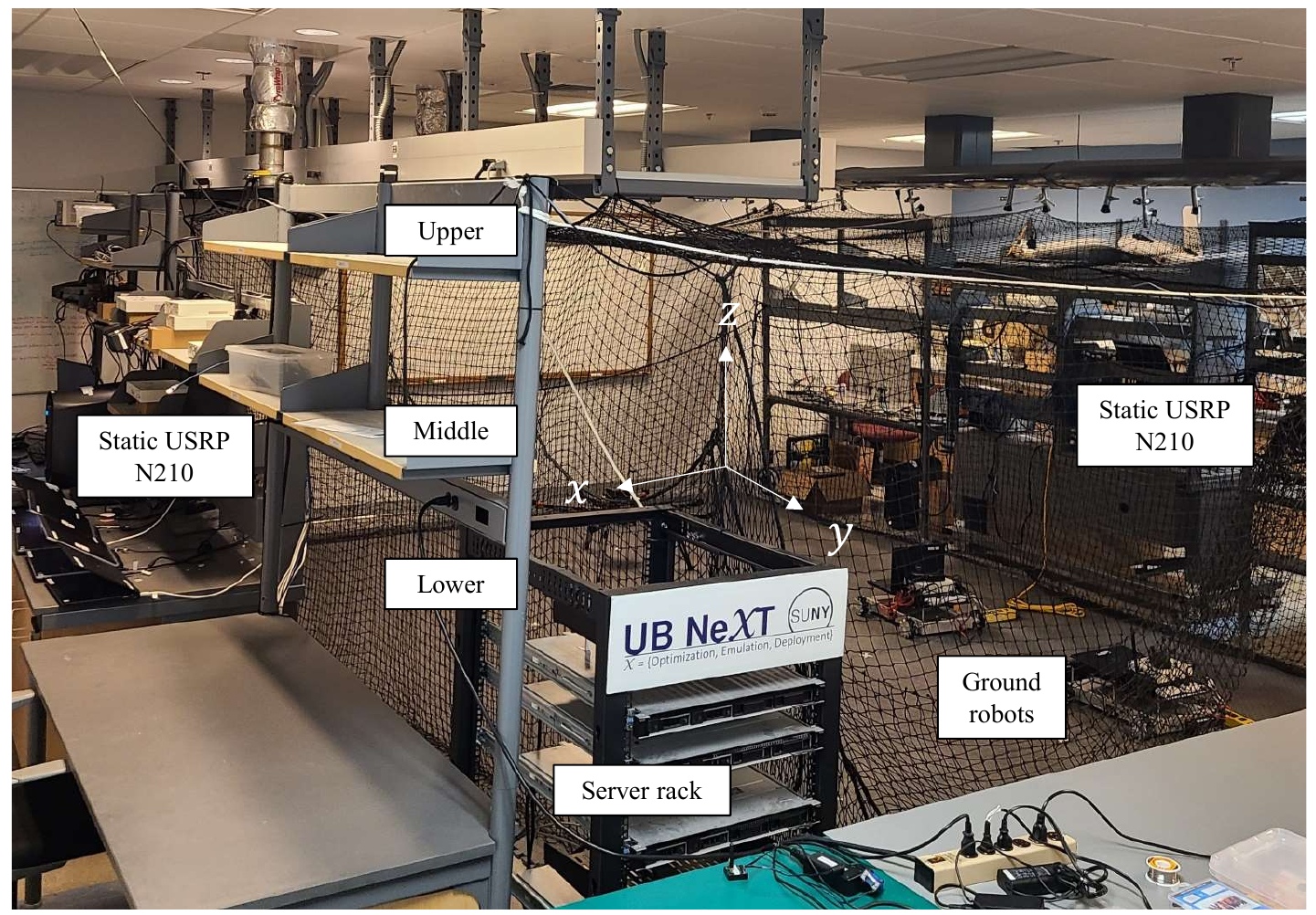} & \includegraphics[width=0.4\textwidth]{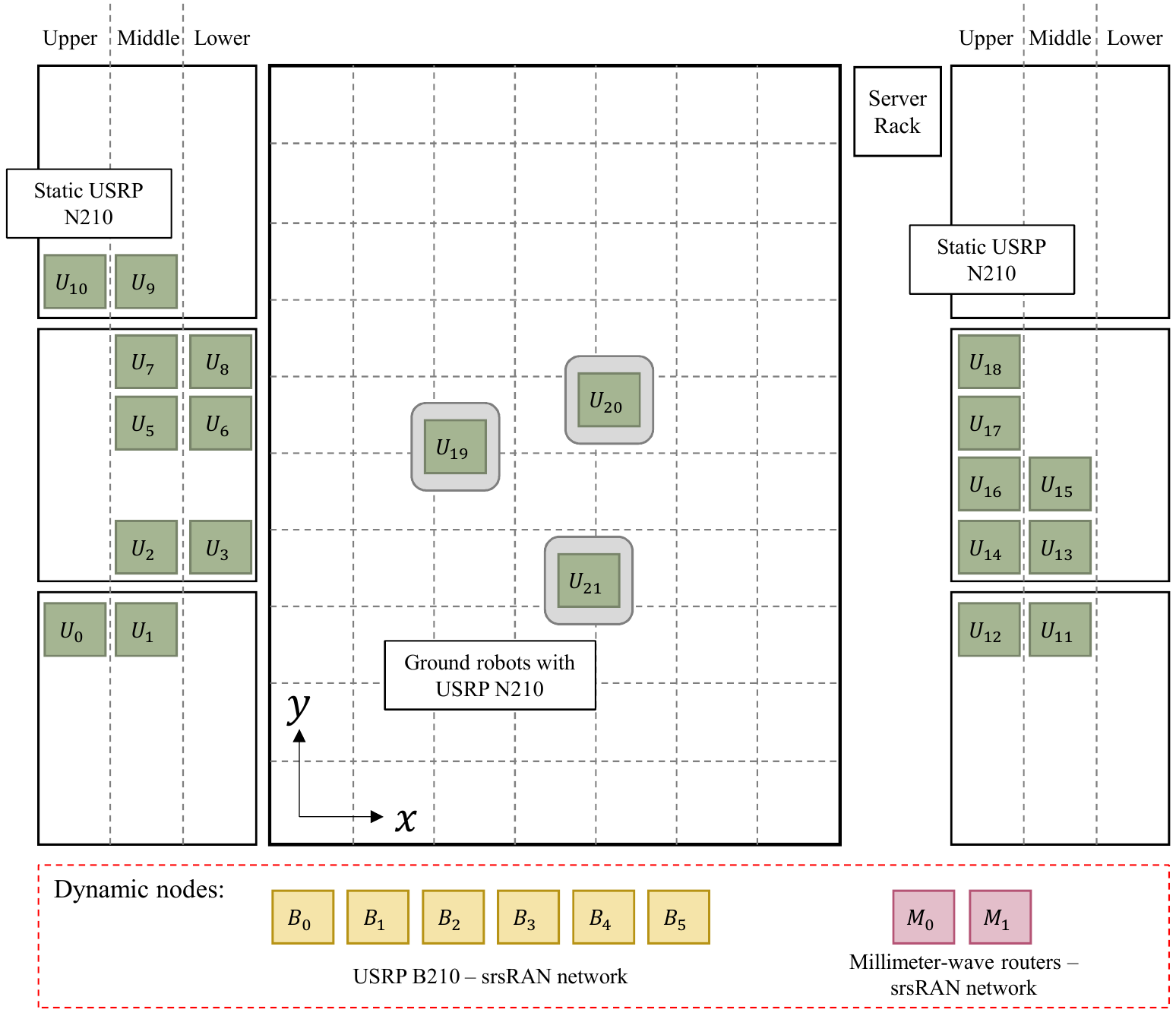} \\
\small (a) &\small \hspace{0mm} (b)\\
\end{tabular}
\caption{\small (a) Snapshot of the UB NeXT testbed; (b) UB NeXT testbed topology.
}
\label{fig:ubnext}
\end{center}
\vspace{-5mm}
\end{figure*}  

\textit{Real-time Training}: 
In ML applications, especially methods that leverage neural networks, the training process is generally time-consuming. 
Even considering faster-than-real-time training capabilities provided by a DT system, significant system or environmental changes may require re-training some or all of a learned policy. In time-critical applications, this may require interim behavioral models to be leveraged during policy re-training. 
The authors of \cite{Changyang2021Tut} explore the integration of low-complexity inference models with ML algorithms to reduce the impact of long training times on system performance in such scenarios. 
Additionally, when simultaneously optimizing multiple agents, as in multi-agent reinforcement learning (MARL), the computation complexity and communication overhead increase exponentially due to the additional problem dimensionality. 
This may further exaggerate the sim-to-real gap based on the aggregate generalizations made across multiple agent representations in the twin domain. 
In the example of a self-coordinating swarm UAV network, each agent may be required to relay significant state information including location, speed, height, and network status of itself and other agents to the twin domain in each training step to avoid collisions, reduce interference, and take actions without loss of information.
Additionally, as the number of agents increase, the amount of 
information required by the twin domain to maintain an accurate model of the physical domain system
increases as well, which further increases network resource consumption. 
The complexity of algorithm design and deployment can be further increased when considering a decentralized scenario, 
in which a twin domain model needs to be maintained at each agent
instead of a central controller.

The Advantage Actor-Critic (A2C) algorithm \cite{Han2019A2C} has been demonstrated as an effective tool to minimize policy training times considering high-dimensional state and action spaces. 
A2C uses the estimated optimal state-action value to update the policy, of which the gradient can be calculated as follows:
\begin{align}
    \nabla_{\theta}J(\theta) = \mathbb{E}_{\pi_{\theta}} [\nabla_{\theta} \log{\pi_{\theta} (s, a)}  A_{\pi} (s, a)]
\end{align}
where $ A_{\pi} (s,a) = Q_{\pi_{\theta}} (s, a) - V_{\pi_{\theta}} (s)$ is termed the \textit{Advantage function}, in which $Q_{\pi_{\theta}} (s, a)$ is the action-value function and $V_{\pi_{\theta}} (s)$ is the state-value function. 
With this advantage function, variance of the gradient can be reduced which improves model training stability. 
It is very time consuming to find hyperparameters that stabilize the learning process, since A2C relies on the initial estimation of values, therefore A2C algorithms can be challenging to design or further time-consuming for real-time training scenarios.

To further accelerate policy convergence, an Asynchronous Advantage Actor-Critic (A3C) \cite{Liu2021A3C} can be used.
Similar to A2C, A3C 
uses an advantage function to reduce variance and improve training stability. The only difference is that A3C allows agents to interact with the environment in parallel. 
In A3C, virtual agents work individually in multiple instances of the same environment to update a global policy asynchronously. This parallelism can significantly reduce policy convergence times. 



To reduce the communication overhead of both centralized and decentralized MARL scenarios, the Lazily Aggregated Policy Gradient (LAPG) method \cite{Chen2022LAPG} can be used to reduce the frequency of communication. 
Most information 
related to 
collision avoidance and 
task completion 
does not
need to be exchanged in every time period, 
e.g.,
sensor malfunctions, low battery states, and obstacle detection. 
LAPG sets a trigger condition for this kind of information to reduce the exchange frequency, which can reduce the overall network communication overhead and computational complexity. The lower bound of the trigger condition for LAPG can be written as:
\begin{align}
\hspace{-3mm}    || \delta \hat{\nabla}^{k}_{m} || ^2 \geq \frac{\xi}{\alpha^2 M^2} \sum_{d = 1}^{D} || \theta^{k+1-d} - \theta^{k-d}||^2 + 6\sigma^2_{m, N, \delta/K},
\end{align}
where $\delta \hat{\nabla}^{k}_{m}$ represents the importance of updating the information, which is calculated by the difference between previous and current policy parameters.

Another strategy that can be used to reduce the per-update communications overhead of a distributed network is by using a 
Partially Observable Markov Decision Process (POMDP) \cite{Lauri2022POMDP}. 
Instead of requiring the agents to fully observe the environment in each time step, action selection for each agent is based on a probability distribution 
given by the model instead of directly 
observing the underlying state. In this case, each agent has less information to 
maintain in the local twin domain model
and the required information to be shared between agents per network update can be reduced.

\section{Physical Scenarios Development}\label{sec:scenarios}

While several works have proposed DT framework concepts to support adoption of DT-enabled technologies at scale \cite{KAZI2017, Huan20, Latif21}, the state of the art in this area generally relies on performance inference based on sim-to-sim experimentation, inflexible virtualization of pre-defined physical scenarios, or small-scale sim-to-real experiments with numerous experimental constraints. 


\subsection{State of the Art} 
Scenario development is a key consideration for high-quality wireless experimentation platforms, which must support user-defined network topologies, protocols, and control problems in order to provide accurate validation for use in a practical DT system. 
The NSF PAWR platforms, including POWDER, COSMOS, AERPAW and ARA, represent the state-of-the-art for wireless network scenario development and experimentation, considering scale, accessibility, and capability. For UAV-enabled wireless networking research, AERPAW \cite{marojevic2020aerpaw} provides a large-scale experimentation platform comprised of static nodes, mobile ground nodes, and UAV systems equipped with SDR hardware. The goal of AERPAW is to 
provide a general platform
to develop and evaluate new capabilities for UAV-enabled wireless networks, and is envisioned to enable research into scalable zero-touch control systems for hybrid aerial-ground networks. 
POWDER \cite{breen2020powder}
is a city-scale wireless networking research testbed in Salt Lake City, Utah, specializing in topics such as 5G O-RAN, massive MIMO, and spectrum sharing in the sub-6 Ghz band. 
COSMOS \cite{rauchaudhari2020cosmos} 
is a testbed deployed in an ultra-dense area of New York City, specializing in research for ultra-high-bandwidth, low-latency wireless communications, millimeter-wave MIMO and beamforming, and advanced edge computing scenarios. 
Finally, ARA 
\cite{zhang22ara}
is a wireless living laboratory focused on enabling research into rural broadband wireless connectivity by connecting an open-access software-defined virtual infrastructure with a heterogeneous mesh of terrestrial radio hardware and LEO satellite communication terminals, capable of providing 600 square miles of contiguous wireless coverage. 


\begin{figure*}[t]
    \centering
    \includegraphics[width=0.85\textwidth]{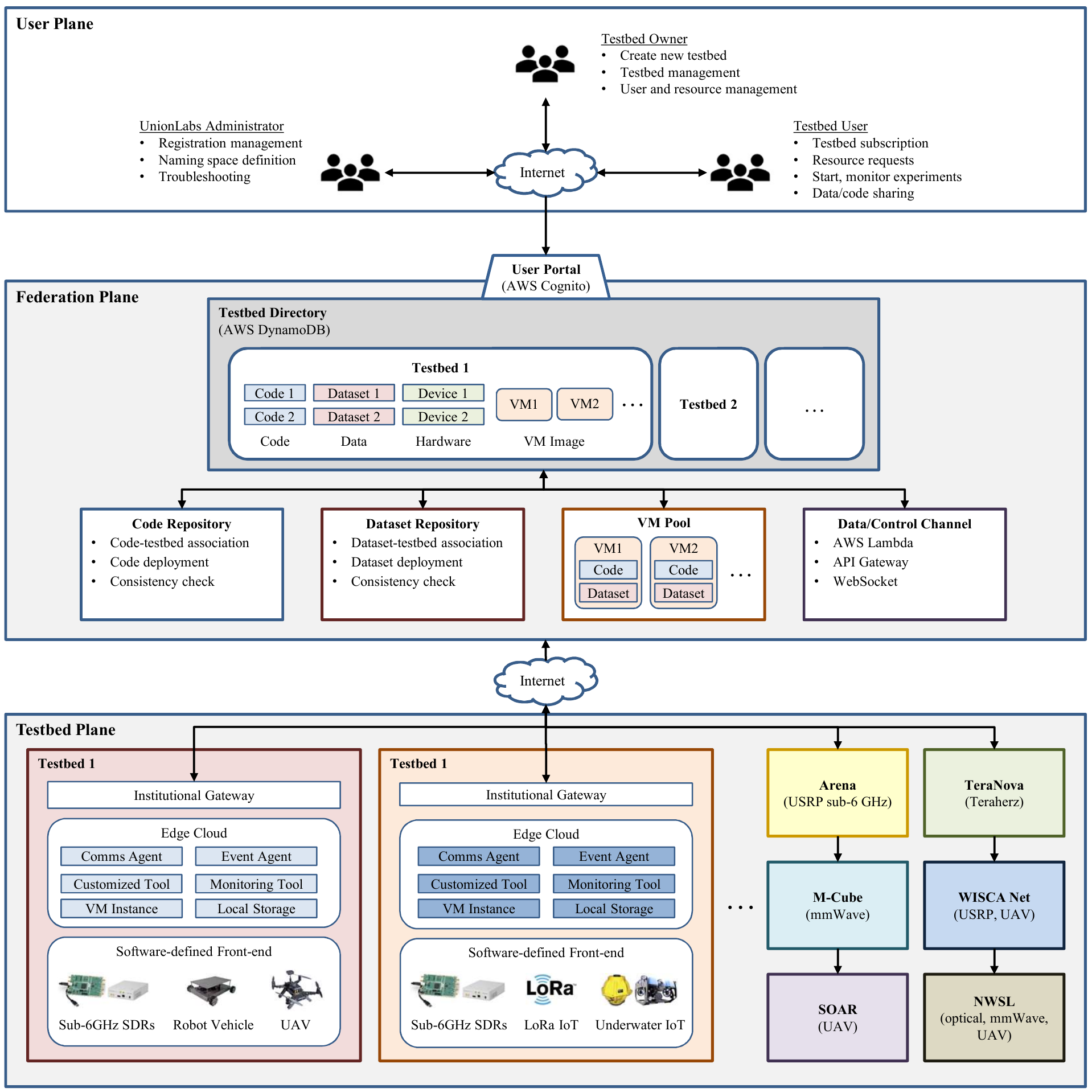}
    \caption{\label{fig:unionlabs}Overview of \textit{UnionLabs} testbed federation.}
\end{figure*}

In addition to AERPAW, the UB NeXT testbed \cite{hu2022ubnext} provides a comprehensive framework for network virtualization and domain adaptation research in integrated aerial-ground wireless networking. We have included a picture and topology diagram of the UB NeXT testbed in Fig.~\ref{fig:ubnext}.
The NeXT testbed platform is part of the UB indoor autonomy research facility, and is comprised of 21 USRP N210 SDRs, 6 USRP B210 SDRs, and two millimeter-wave routers, with mobility support provided by three ground robots with 22 kg payload capacities and a netted UAV enclosure for safe aerial network testing. The testbed networking environment has been fully virtualized
in UBSim, which has been demonstrated in \cite{mcmanus2022sourcetotarget}. This testbed can enable rapid, small-scale experimentation to address existing challenges in DT research such as evaluation of the \textit{sim-to-real} gap, integrated optimization-learning algorithm design for efficient ML, and virtual network self-configuration via system identification. 



\subsection{Research Opportunities}

Scenario development is at the core of validating DT-enabled experimental frameworks for the wireless domain. We identify several key research opportunities for expanding the scope and depth of continued research in this direction. 

\textit{Sim-to-real gap Estimation}: 
In general, domain adaptation methods seek to bridge the gap between physical and DT domains. However, especially in the case of robust learning, estimation of the \textit{sim-to-real} gap may not guarantee optimal performance if the gap between physical and DT domain behaviors is large or unknown. 
Furthermore, since there is no unifying framework for \textit{sim-to-real} gap measurement, methods that seek to reduce the \textit{sim-to-real} gap may require manual tuning in the case of multi-physics optimization. 
System identification has shown promise in reducing the performance gap between physical and DT domains for physical scenarios regarding mechanical or robotic systems \cite{jiang2021simgan}, but 
there is insufficient investigation into how to quantify the \textit{sim-to-real} gap between different physical scenarios for other methods of domain adaptation. 
Further research into the measurement or estimation of the \textit{sim-to-real} gap induced by various physical networking scenarios is anticipated to accelerate design of domain adaptation schemes and hence advance the state-of-the-art of practical DT-enabled wireless networking. 

\textit{Portable environments for UBSim}: We demonstrate in \cite{mcmanus2022sourcetotarget, sabarish2022flytera} the need for multiple environmental models to enable experimentation in domain adaptation, with specific attention to both sim-to-sim and sim-to-real gaps. Specifically, more virtual models will be made available for future work to build on the contributions in \cite{mcmanus2022sourcetotarget}, enabling rapid and repeatable experimentation for domain adaptation in the wireless domain through sim-to-sim experimentation. By virtualizing real testbeds, as done in \cite{mcmanus2022sourcetotarget}, this will provide preliminary benchmark results required to motivate continued research for sim-to-real transfer. 

Building on the sim-to-sim framework outlined in \cite{mcmanus2022sourcetotarget}, we identify the need for a flexible sim-to-real domain adaptation framework that can accommodate different environmental models based on the physical domain specification. This will enable the design of new domain adaptation algorithms for the wireless domain as well as adaptation of existing algorithms. Using the same simulation platform as \cite{mcmanus2022sourcetotarget} and \cite{sabarish2022flytera}, as well as the indoor autonomy research facility and UB SOAR facility at University at Buffalo, many network configurations can be observed, including heterogeneous aerial-ground networks and UAV-to-UAV networks.
To achieve the short-term goal of sim-to-real experimentation, we plan direct integration with the UB NeXT testbed platform \cite{hu2022ubnext}. In preliminary sim-to-sim experiments, we have virtualized the NeXT testbed \cite{mcmanus2022sourcetotarget} and will use this DT environment to better understand the sim-to-real gap through rigorous sim-to-real experimentation and domain adaptation algorithm design. 


\textit{Testbed Sharing and Remote Access}: 
Considering the need for open, accessible DT experimentation platforms, we believe it is of critical importance to facilitate remote access and control for a fully realized DT-enabled wireless networking testbed. 
There remains a lack of testbeds and networking environments to enable validation of AI integration and further research into virtualization for network autonomy \cite{coronado2022zerotouch}, especially to support advancement towards zero-touch networking. 
To address this challenge, we emphasize the contributions made in \cite{moorthy2022cloudraft} as discussed in Sec.~\ref{sec:virtualization}, and propose an expansion of the supported framework to include simulation/emulation capabilities.
We envision a new framework referred to as \textit{UnionLabs} for testbed sharing and federation. As illustrated in Fig.~\ref{fig:unionlabs}, the architecture of UnionLabs consists of three planes, connected by the internet: the \textit{User Plane}, which handles user/operator interactivity, registration, and management; the \textit{Federation Plane}, which coordinates testbed access and stores experimental code, datasets, and virtual machines; and the \textit{Testbed Plane}, which is comprised of all federated testbeds connected through institutional gateways.  This initiative will provide a platform to share code, data, and software/hardware resources across a federation of cloud-enabled heterogeneous testbeds distributed throughout the country, with an emphasis on the advancement of research topics related to NextG wireless networks, zero-touch and network automation, and the wireless Internet of Things. 

\section{Conclusions}\label{sec:conclusion}
In this work, we reviewed existing literature regarding the use of DTs for ML-enabled wireless networks with an emphasis on UAV-enabled networking, and discussed the open research challenges in the area. 
DT for the wireless domain is a particularly important open research area, and can serve as an enabling technology for practical applications of data-driven network self-optimization such as UAV network self-coordination and autonomous network control. 
Domain adaptation, a key element to bridge the gap between simulations and real network deployments, requires further investigation in the wireless domain. 
Several methods of domain adaptation, including system identification, domain-agnostic feature extraction, and robust learning have been evaluated for use in a DT system, focusing on limiting interactions between domains to improve data efficiency. However, a comprehensive exploration of the reality gap present in DTs remains an open challenge in this area, as well as accelerating real-time training using domain adaptation in multi-agent systems such as UAV swarm networks. 
In order to further identify the reality gap across domains, the topic of physical scenario development also needs to be further explored especially for wireless UAV networks. 

\bibliographystyle{ieeetr}
\bibliography{survey}
\end{document}